\documentclass{ws-ijgmmp}

\begin{document}

\markboth{Authors' Names}
{Instructions for Typing Manuscripts (Paper's Title)}

%
\catchline{}{}{}{}{}
%

\title{Fold bifurcation entangled surfaces for one-dimensional Kitaev lattice model}

\author{Davood Momeni}

\address{Department of Physics, College of Science, \\Sultan Qaboos University Al Khodh 123 , Muscat, Oman\\
davood@squ.edu.om}

\author{Phongpichit Channuie}

\address{College of Graduate Studies, \\Walailak University, Thasala, Nakhon Si Thammarat, 80160, Thailand\\
School of Science, Walailak University, Thasala, \\Nakhon Si Thammarat, 80160, Thailand\\
Research Group in Applied, Computational and Theoretical Science (ACTS), \\Walailak University, Thasala, Nakhon Si Thammarat, 80160, Thailand\\
channuie@gmail.com}

\maketitle

\begin{history}
\received{(Day Month Year)}
\revised{(Day Month Year)}
\end{history}

\begin{abstract}
We investigate a feasible holography with the Kitaev model using dilatonic gravity in AdS$_2$. We propose a generic dual theory of gravity in the AdS$_2$ and suggest that this bulk action is a suitable toy model in studying quantum mechanics in Kitaev model using gauge/gravity duality. This gives a possible equivalent description for the Kitaev model in the dual gravity bulk. Scalar and tensor perturbations are investigated in details. In the case of near AdS perturbation, we show that the geometry still "freezes" as is AdS, while the dilation perturbation decays at the AdS boundary safely. The time-dependent part of the perturbation is an oscillatory model. We discover that the dual gravity induces an effective and renormalizable quantum action. The entanglement entropy for bulk theory is computed using extremal surfaces. We prove that these surfaces have a fold bifurcation regime of criticality.
\end{abstract}


\section{Introduction}
Information is the building block of the whole physical universe and it is mainly believed that it can explain the origin of the thermodynamics, specially in the black holes, in the context of the general theory of relativity \cite{z12j,jz12}. Another commonly accepted idea is that, certain types of the geometries for the spacetime may emerge from the quantum system, mainly from the entanglement between quantum states, and it is motivated mainly from the celebrated holographic principle, where the information of a region of spacetime can be decoded through its boundary surface \cite{1, 2}. A complete description of the holographic principle is physically possible if this holographic conjecture is interpreted as a gauge/gravity duality between certain types of the quantum field theories with conformal symmetry and specific geometries with negative curvature. This is a definition of the AdS/CFT correspondence~\cite{M:1997}. As an attempt to deeper understand this conjecture, we has been trying to explain quantum entanglement based on the holographic principle in a consistent form. Note here that in the gauge/gravity duality, the bulk theory has one dimension more than the specific boundary quantum theory. In a limited case via Maldacena's conjecture the string commodification ends to a dual CFT. As an example the AdS$_5$/CFT$_4$ is a meaningful valid duality, although it just remains as a conjecture without proof. We would like to mention that the original superconducting quantum wire model of Kitaev is trivially a $0+1$ dimensional model as the standard STK model. If we can suggest any bulk dual for it, naturally it must have $1+1$ dimensions. This clarification helps the readers to see the logical connection and application of the AdS/CFT for Kitaev model.

In condensed matter physics it has always been illustrative to figure out gravitational dual models which are able to extract the same (almost exact) physical results for the system using a simpler method. One of the most successful examples in condensed matter is the unpaired Majorana fermions and we already knew that it should correspond to superconductive systems. A physical realization of such systems was made by Kitaev \cite{kitaev}, when he considered a quantum wire lying on the surface of 3-dimensional superconductor. If we consider a chain consisting of large numbers of possible quantum sites, each quantum site can be either occupied or unoccupied by an electron by fixing the direction of each electron.

In this work we will investigate the dual properties of a  Kitaev model through the use of the gauge/gravity duality in two dimensional AdS spacetime. Our aim is to deeper understand the physical features of the conformal geometry and the chaos in AdS$_2$ duals of a Kitaev system in a general way without making assumptions of conformal invariance in the original Kitaev model. To this end we studied the entanglement properties of the dual gravity model using  Ryu-Takayanagi approach. 

The organization of the paper is as the following: 
In Sec.\ref{1DKitaev}, we explain a framework of a one dimensional Kitaev quantum wire model. 
In Sec.\ref{dualtoy}, we propose a toy model for a Kitaev model using gauge/gravity duality. 
In Sec.\ref{spt}, we treat the scalar perturbation of the system. 
In Sec.\ref{adspt}, we investigate near AdS geometry.
In Sec.\ref{Holo}, the holographic renormalization is investigated.
In Sec.\ref{entang}, the Kitaev entanglement entropy is
computed via holography. In Sec.\ref{fold}, we first propose the idea of fold bifurcated entangled surfaces as a root for chaos. Finally, we conclude our findings in the last section.

\section{One dimensional Kitaev quantum wire model}
\label{1DKitaev}
The one dimensional Kitaev quantum wire as a one dimensional lattice model for a superconductor is perfectly described by the following Hamiltonian~\cite{kitaev}:
\begin{eqnarray}\label{Hkitaev}
H=  - \eta \sum^N_{j=1}  (c_j^\dagger c_{j+1}+c_{j+1}^\dagger c_j)
-\mu\sum^N_{j=1}  (c_j^\dagger c_j-{\textstyle\frac{1}{2}})+\sum^N_{j=1}  ( \Delta c_jc_{j+1}+\Delta^* c_{j+1}^\dagger c_j^\dagger ),
\end{eqnarray}
where $c_j^\dagger$ represents creation operator for the electron, $\eta$ is called the hopping integral, $\mu$ is the chemical potential, $\Delta\in\mathcal{C}$ is the superconducting gap of energy, and $N$ is the lattice size. Phase transition in Kitaev model happens when we adjust the model parameters with $\mu=2\eta$ for $|\Delta|>0$, where the the energy spectrum of the system is given by 
\begin{eqnarray}
E_k= \sqrt{(2\eta \cos k + \mu)^2 + 4 |\Delta|^2 \sin^2 k}.\label{Ek}
\end{eqnarray} 
Recently it has been shown that the type of phase transition in Kitaev model is a topological quantum phase
transition  and it can
be characterized by nonlocal-string order parameters \cite{phasetransition}. Choosing the chemical potential $\mu$ as the external variational  parameter, one can show that the quantum distance between two nearby ground state wave functions of the system (what is known as the fidelity in the literature) can be computed easily 
and this quantity can be used as an order parameter to address certain types of the phase transitions  in strongly coupled systems \cite{XFS}.

The Kitaev model and its generalizations always illustrate several phase transitions as well as chaotic behavior. There are many interesting works related to chaos simply by random quantum mechanical systems. Regarding the existing references, we can discuss some more interesting results published so far in the literature. In a quantum many-body system, an exactly solvable model, now known as the complex SYK model, was initially proposed by Mon and French. An earlier slightly different version of the model appeared in Refs.\cite{French1970,French1971,Mon1975} and was applied to describe the spectral properties of nuclei. A mean field solution of this model was obtained by Sachdev and Ye and Kitaev formulated this model for Majorana fermions. There have been many different aspects of this novel theory investigated by researchers. Remarkable observations about the SYK models are in the fact that they display the existence of disordering and chaos. For example in \cite{Liu:2016rdi}, the mesoscopic features of the SYK model were investigated for $q\gg \sqrt{N}$ and mainly it was shown that how the dynamics obeys a stochastic dynamics and consequently the time evolutions can be described in a dual form in terms of random matrix theory (RMT). It is worth mentioning that the SYK model becomes an RMT model for large $q$ was well known before, even to Mon and French. In the opposite regime when $q\ll \sqrt{N}$, the system shows a high diffusion at the early chaotic phase. It is worth mentioning that the analysis for $q \gg \sqrt{N}$ was understood much earlier in work by Benet and Weidenmuller \cite{Benet:2002br}. 
More works on the RMT have been done during the ground breaking studies, e.g. \cite{Benet2},\cite{Benet3}. Chaos in the SYK model was also related to the criticality in the Fermi surfaces in super fluidity phase \cite{Patel:2016wdy}.  The parameters related to a toy many body theory dual to SYK was examined near the criticality at nonzero density. To have symmetry breaking an auxiliary $U(1)$ was introduced, and to demonstrate the chaotic behavior of the system the authors computed Lyapunov exponents. Here an extended velocity in the phase portrait was found (the butterfly velocity). Notice that a replica path integral approach describing the quantum chaotic dynamics of the SYK model at large time scales was also studied in Ref.\cite{Altland:2017eao}.

It was demonstrated numerically that the thermal parameters have a universal behavior and parametrize the chaos in an extended phase space. The regime of chaos was limited  only to the high temperature systems although it is possible to have chaos in enough low-finite critical temperature ranges. In Ref.\cite{Gu:2016oyy}, a possible local criticality was investigated instead of global ones. This recent work generalizes $0+1$ dimensional SYK model to higher dimensions as an attempt to study further criticality aspects. To quantify the chaos as disordering in the propagation of the perturbations, a butterfly velocity was introduced with finite velocity and it can characterize the chaotic effect in the system. However, the quantum chaos seems not to be limited to SYK model. For example tensor models also have chaotic dynamics for example because of very large numbers of degrees of freedom \cite{Krishnan:2016bvg}. Almost all these references were satisfied the SYK model from the dual quantum theory point of view.  Till now there is no work devoted to study chaos from the bulk theory, specially the chaotic and phase transitions from the dual bulk geometry point of view. If one can construct a bulk model for the original  model defined by the Hamiltonian (instead of the Lagrangian which is needed to build bulk theory as a dual gravitational theory) presented in the Eq. (\ref{Hkitaev}), we can probe the possibility to have chaos in the geometry; for example chaotic minimal surfaces (dual to the chaotic entanglement entropy), to address chaos in the quantum boundary theory. 

Our motivation is to use the important work of Ryu-Takayanagi to define entanglement entropy via AdS/CFT proposal. The AdS/CFT correspondence can provide a technique to find the holographic entanglement entropy (HEE) of a quantum system using minimal surfaces in the dual bulk gravity theory. This technique was discovered by  Ryu-Takayanagi, for any arbitrary subsystem $A$ in the quantum boundary theory, we need to specify the $(d-1)$-minimal surface which is extended into the AdS bulk with the same boundary with $A$ (the boundary is denoted by $\partial A$). The HEE for this subsystem via the celebrated principle of Ryu-Takayanagi can be written as \cite{hee1},
\begin{eqnarray}
S_{\rm HEE}=\frac{{\rm Area}(\gamma_A)}{4G}\label{hee}
\end{eqnarray}
where $G$ is the gravitational coupling  in the AdS spacetime and $\gamma_A$ is the minimal area surface. In this paper we show how that the minimal surfaces corresponding to the entanglement entropy in the dual picture have chaotic form. 

\section{Building a dual toy model}
\label{dualtoy}
In this section we aim to propose a toy model for Kitaev model using gauge/gravity duality, where we will be able to address phase transitions using gravity dual model in bulk. According to the AdS/CFT, any deformation of the boundary theory operators can make a change in the bulk theory either in terms of the bosonic or ferminoc or even other gauge forms. What we suggested here deserved to be considered as a viable dual bulk for a theory used by Kitaev.
To build up a holographic model we must move from Fock space  to a continuous bulk model, with matter fields. A possible set of physical fields is a set of scalar fields $\phi$. It was shown that in Kitaev model with the periodic boundary condition the Fourier transformation is can be implemented, where momentum $k$ is a good quantum number \cite{XFS}. In the Fock space, a pair of electrons with momentum $k$ and $-k$ will form the Cooper pair and the superconducting phase can exist.

There is no any simple method to build the toy model bulk theory for the given Kitaev Hamiltonian, given in Eq.(\ref{Hkitaev}). For the SYK model, the model just proposed as an ad-hoc dilatonic action in two dimensional background, posses an AdS black hole as bulk geometry \cite{prl}. Although both SYK and Kitaev theories are defined using a pair of creation and annihilation operators for Dirac particles and it seems that the dual bulk model should be written in the form $S[g,\Psi,...]$ (here $g$ is the metric, $\psi$ is doublet fermionic vector), but surprisingly it is always possible to build a bosonic bulk sector for this fermionic boundary action. This may be probably connected by mean of supersymmetry or any other hidden superalgebra. However, we are not going to deal with the symmetric aspects of the model. The aim of this section is to explain how a simple finite sized bosonic lattice can be dual to the non relativistic, non conformally symmetric model presented in Eq.(\ref{Hkitaev}). Note that the Hamiltonian is written in the Fock space. If we suppose that the Fock space describes a continuous spanned Fourier modes space, then we label to each operator $c_j$ with a dual bulk bosonic field $\phi$ located at the position of a hypothetical lattice $\vec x$. The aim is to build enough invariant bulk model. Since we don't have conformal symmetry in the original boundary quantum theory, we will define the bulk as symmetrically as the Lorentz symmetry is preserved. Consequently, the lattice location corresponding to the operator $c_j$ is $\vec x=(t,r)$ ($c=1$). Note that the bosonic field corresponding to the closest neighbor of the first is $\phi(\vec x-\vec a)$ for $c_{j+1}$ and we can define the conjugate field $\bar \phi$ for $c_j^{\dagger}$. The basic equations for our proposed duality are given as following:
\begin{table}[ht]
\caption{The table shows a possible duality between Kitaev theory as boundary and a type of doublet scalar dilaton in AdS bulk background. The elements of the boundary action are presented as a pair of fermionic operators while the bulk is a pair of doublet scalar fields, dilatonic fields. To define dual operators we use a lattice scale $a$ dual to an UV cutoff in boundary proportional to $E_{UV}\propto a^{-1}$.}
\begin{center}
\label{hubbleq}
\begin{tabular}{cccccccccc}
\hline\hline
&{\rm Fock\,space}& & & &  & & & & {\rm Bulk\,geometry}\\
\hline\hline
&$c_j$& &  & &  &  & & & $\phi(\vec x)$\\
&\,\,\,\,\,\,$c_{j+1}$ & &  & &  & & & &\,\,\,\,\,\,\,\,\,\,$\phi(\vec x+\vec a)$\\
&$c_j^{\dagger}$ & & & &  & & & & $\overline{\phi}(\vec x)$\\
&\,\,\,\,\,\,$c_{j+1}^{\dagger}$ & & & &  & & & & $\,\,\,\,\,\,\,\,\,\,\overline{\phi}(\vec x+\vec a)$\\
\hline\hline
\end{tabular}
\end{center}
\end{table}

Note that the right-hand side quantities are defined in the bulk geometry, while the left-hand side is defined in the Fock space. Now consider the vector $\vec a$ (the vector of lattice) having infinitesimal number (magnitude) and corresponding to an ultraviolet (UV) cutoff of order $a^{-1}$. If we use the Taylor series in the vector form, i.e., 
\begin{eqnarray}
\phi(\vec x+\vec a)=\sum_{n=0}^{\infty}\frac{(\vec a\cdot \nabla)^n \phi(\vec x)}{n!},\label{series}
\end{eqnarray}
keeping the terms up to the order $\mathcal{O}(a)$ we can rewrite the Kitaev Hamiltonian in a familiar form assuming the limit of $|a|\to 0$ and $\sum\to \int d\rho(\vec x)$, where $d\rho$ is a length measure in bulk theory. We substitute the above analogies into the Hamiltonian $H$ given in Eq.(\ref{Hkitaev}) and then obtain
\begin{eqnarray}\label{Hfock}
{\cal H}&=& - \eta \Big[\overline{\phi}(\vec x)\phi(\vec x+\vec a)+\overline{\phi}(\vec x+\vec a)\phi(\vec x)\Big]
-\mu\Big(\overline{\phi}(\vec x)\phi(\vec x)-{\textstyle\frac{1}{2}}\Big)\nonumber\\&&\quad+\Delta \phi(\vec x)\phi(\vec x+\vec a)+\Delta^* \overline{\phi}(\vec x+\vec a)\overline{\phi}(\vec x),
\end{eqnarray}
where $H=\int d\rho(\vec x) {\cal H}$. Invoking Taylor series up to ${\cal O}(a)$ given in Eq.(\ref{series}), the above expression becomes
\begin{eqnarray}\label{Hfockdd}
{\cal H}&\approx& - \eta \Big[\overline{\phi}(\vec x)\Big(\phi(\vec x)+(\vec a\cdot \vec \nabla)\phi(\vec x)\Big)+\Big(\overline{\phi}(\vec x)+(\vec a\cdot \vec \nabla)^{\dagger}\overline{\phi}(\vec x)\Big)\phi(\vec x)\Big]
-\mu\Big(\overline{\phi}(\vec x)\phi(\vec x)-{\textstyle\frac{1}{2}}\Big)\nonumber\\&&\quad+\Delta \phi(\vec x)\Big(\phi(\vec x)+(\vec a\cdot \vec \nabla)\phi(\vec x)\Big)+\Delta^* \Big(\overline{\phi}(\vec x)+(\vec a\cdot \vec \nabla)^{\dagger}\overline{\phi}(\vec x)\Big)\overline{\phi}(\vec x).
\end{eqnarray}
Also, note that the total action must be a real valued function, i.e. $\phi=\overline{\phi}$. In a more compact form, we write
\begin{eqnarray}\label{Hfockre}
{\cal H}&\approx& \overline{\phi}(\vec x)\phi(\vec x)(-2\eta-\mu) +\Delta \phi(\vec x)^{2} +\Delta^{*}\overline{\phi}(\vec x)^{2}+(\vec a\cdot \vec \nabla)\big[-\eta \overline{\phi}(\vec x)\phi(\vec x)+\Delta \phi(\vec x)^{2}\big]\nonumber\\&&+(\vec a\cdot \vec \nabla)^{\dagger}\big[-\eta \overline{\phi}(\vec x)\phi(\vec x)+\Delta^{*} \overline{\phi}(\vec x)^{2}\big]+\frac{\mu}{2}.
\end{eqnarray}
However, if we limit this work to the dilaton profile, i.e. $\phi=\overline{\phi}$, we obtain
\begin{eqnarray}\label{Hfockredi}
{\cal H}\approx\phi(\vec x)^{2}\big[-2\eta-\mu +\Delta +\Delta^{*}\big]+\phi(\vec x)\lambda^{\nu}\nabla_{\nu}\phi(\vec x)\big[-2\eta +\Delta +\Delta^{*}\big]+\frac{\mu}{2},
\end{eqnarray}
where we have replaced $(\vec a\cdot \vec \nabla)\phi$ with $\lambda^{\nu}\nabla_{\nu}\phi$ and $\mu$ is just a constant. The last expression allows us to determine the Lagrangian. Since the dilatonic action is defined as $\int d^2x\sqrt{-g}(\phi R)/2\kappa^2$, hence we multiply (\ref{Hfockredi}) by a factor $-1/2$ to yield
\begin{eqnarray}\label{Hfockredim}
{\cal H}\longrightarrow\phi(\vec x)^{2}\Big[\eta +\frac{\mu}{2} -\frac{\Delta +\Delta^{*}}{2}\Big]+\phi(\vec x)\lambda^{\nu}\nabla_{\nu}\phi(\vec x)\big[\eta -\frac{\Delta +\Delta^{*}}{2}\big]-\frac{\mu}{2},
\end{eqnarray}
where the last term is just a constant. Note here that the conjugate momentum is $\Pi_{\mu}\propto \nabla_{\nu}\phi$. Then the Lagrangian density takes the form
\begin{eqnarray}\label{Lfockredim}
{\cal L}&\longrightarrow&\phi(\vec x)^{2}\Big[\eta +\frac{\mu}{2} -\frac{\Delta +\Delta^{*}}{2}\Big]-\phi(\vec x)\lambda^{\nu}\nabla_{\nu}\phi(\vec x)\big[\eta -\frac{\Delta +\Delta^{*}}{2}\big]-\frac{\mu}{2}\nonumber\\&=&\phi(\vec x)^{2}\Big[\eta +\frac{\mu}{2} -\frac{\Delta +\Delta^{*}}{2}\Big]+\phi(\vec x)\lambda^{\nu}\nabla_{\nu}\phi(\vec x)\big[\frac{\Delta +\Delta^{*}}{2} - \eta \big]-\frac{\mu}{2}.
\end{eqnarray}
We note that $\mu/2$ is just a constant term in the potential function assumed in the first term, while the kinetic energy is given by the second term. Furthermore to make dynamics of the background we use the minimal coupling of the scalar field, called dilaton, to the scalar curvature $R$:
\begin{eqnarray}
S=\int d^2x\sqrt{-g}\Big(\frac{\phi R}{2\kappa^2}+\theta\phi\lambda^{\mu}\nabla_{\mu}\phi+\beta\phi^2-\frac{\mu}{2}\Big)+S_{\rm matter},\label{S2}
\end{eqnarray}
where $\kappa^2=8\pi G$ denotes the gravitational coupling, $G$ is a gravitational constant, and $R$ is Ricci scalar of the two-dimensional metric $g_{\mu\nu}$. The covariant form for the connections are given by $\nabla_{\mu}=(\partial_t,\partial_r)$, $\mu=\{1,2\}$. A pair of parameters for our toy model is given by 
\begin{eqnarray}
&&\theta=\frac{\Delta+\Delta^{*}}{2}-\eta,\\&&
\beta=\eta+\frac{\mu}{2}-\frac{\Delta+\Delta^{*}}{2}.
\end{eqnarray}
Here $S_{\mbox{matter}}$ is a matter action and $\lambda$ is a Lagrange multiplier. The other parameters like $\Delta,\eta$ are defined in a similar manner as they were defined in the Kitaev model. The reason to introduce a dilaton is that the compactification of the Einstein-Hilbert action in two dimensional Riemannian manifolds generates a non zero dilaton field. The equations of motion for (\ref{S2}) read
\begin{eqnarray}
&&\frac{R}{2\kappa^2}+2\beta\phi +\theta \lambda^{\mu}\nabla_{\mu}\phi =J,\label{eomphi}\\
\label{eomg}&&-\nabla_{\mu}\nabla_{\nu}\phi+g_{\mu\nu}\nabla_{\alpha}\nabla^{\alpha}\phi-\frac{g_{\mu\nu}}{2}
\Big(\beta\phi^2-\frac{\mu}{2}\Big)+\theta\phi\lambda_{\mu}\nabla_{\nu}\phi
=T_{\mu\nu},
\end{eqnarray}
where $T_{\mu\nu},J$ are the energy-momentum tensor and dilaton source given by
\begin{eqnarray}
T_{\mu\nu}=\frac{2\kappa^2}{\sqrt{-g}}\frac{\delta S_{\rm matter}}{\delta g^{\mu\nu}},\ \ J=-\frac{2\kappa^2}{\sqrt{-g}}\frac{\delta S_{\rm matter}}{\delta \phi}.
\end{eqnarray}
We can eliminate $\lambda$ using variation of the action (\ref{S2}). Then by the variation over $\lambda$, and because the variation is given by $\frac{\delta S}{\delta \phi}-\nabla_{\mu}\big(\frac{\delta S}{\delta (\nabla_{\mu}\phi)}\big)$, we obtain the equation of motion for dilaton field $\phi$ given in (\ref{eomphi}).

From dilatonic  equation of motion (\ref{eomphi}), we observe that dilaton toy model which is presented here has AdS$_2$ vacua solution at $\phi=\phi_c$ and $R_c=-\kappa^2\sqrt{\frac{\beta \mu}{2}}$ in the absence of any matter fields and in the AdS$_2$ geometry. These equations take the form
\begin{eqnarray}
\phi=\phi_c,\ \ g=L^2(-r^2dt^2+2dtdr).\label{ads2}
\end{eqnarray}
Note that AdS radius is defined here as an effective radius. Then by solving Eq.(\ref{eomphi}) with respect to $L$, one can find 
$$
L^2=\frac{2}{\kappa^2}\sqrt{\frac{2}{\mu}}.
$$
We take $R_c=\frac{-2}{L^2}=-2$ hereafter. We now work in the infalling Eddington-Finkelstein coordinate  with null coordinates. Using holographic renormalization scheme in the AdS$_2$ throat, we can show that \cite{Grumiller}:
\begin{itemize}
\item It is not possible to make a dual quantum operator for dilaton field. This is because there is no any type of \textsc{one to one} map between different copies of AdS and dual quantum theories.
\item Because the boundary metric at $r\to\infty$ becomes singular, we conclude that the boundary energy-momentum tensor should vanish to preserve conformal symmetry.
\item The algebraic structure for the  dual boundary quantum theory is a Virasoro
symmetry. It has been argued that the central charge for such invariant dual theory is zero \cite{Strominger}. 
\end{itemize}

The dual boundary quantum theory lives at a singular one-dimensional Lorentzian metric where the null path is trivial given by $t=0$ where $t$ is time coordinate. There is an evidence here regarding inconsistency between two-dimensional version of AdS and finite energy excitations in the dual quantum theory and a direct proof is to show that boundary stress tensor vanishes \cite{Maldacena}. Note that in our toy model this statement has been directly implemented because energy momentum tensor for boundary is identically zero.

There is an alternative way to write a dilaton action~(\ref{S2}) in a form of modified gravity. Here we substitute the equation of motion (\ref{eomphi}) in (\ref{S2}). After simple algebraic manipulations, we obtain
\begin{eqnarray}
S=\int\frac{\sqrt{-g}d^2x}{2\kappa_{\rm eff}^2}\Big(-2\Lambda_{\rm eff}-\beta R^2+\theta\lambda^{\mu}R\nabla_{\mu}R\Big)+S_{\rm matter},\label{mog}
\end{eqnarray}
where we have define an effective Newton and cosmological constants (dual to chemical potential $\mu$) and these are given as follow:
\begin{eqnarray}
&&G_{\rm eff}=G\sqrt{2}\big(\Delta+\Delta^{*}-\mu-2\eta\big),\\&&
\Lambda_{\rm eff}=\mu\Big(8\pi G(2\eta+\mu-(\Delta+\Delta^{*}))\Big)^2.
\end{eqnarray}
This yields a new type of modified gravity in two dimensions proposed recently in cosmology~\cite{Tsujikawa} and defines the model in the geometric frame for the action integral (\ref{mog}).

It will be interesting to study black hole solutions in this type of gravity as well as investigating the possibility to have AdS black hole solutions, with $R<0$. The EoM for (\ref{mog}) is written in the following form:
\begin{eqnarray}
&&\theta \lambda^{\mu}\nabla_{\mu}R -2\beta R =0\label{phieom},\\&&
\theta \lambda_{\nu}R\nabla_{\mu}R-g_{\mu\nu}\Big(2\Lambda_{\rm eff}+\beta R^2\Big)=\kappa_{\rm eff}^2T_{\mu\nu}\label{geom}.
\end{eqnarray}
with
\begin{eqnarray}
T_{\mu\nu}=\frac{2}{\sqrt{-g}}\frac{\delta S_{\rm matter}}{\delta g^{\mu\nu}}.
\end{eqnarray}
The existence of an AdS black hole is a crucial feature of gauge/gravity picture. Since in our model we built a modified gravity action equivalent model, we can examine the constant Ricci scalar black hole solutions using the set of the EoMs presented in  Eqs.(\ref{phieom},\ref{geom}). Basically we suppose that there should exist a constant dilaton field $\phi=\phi_c$ which corresponds to a constant Ricci scalar black hole $R=R_c$. In the absence of any matter fields (the matter field will be added for renormalization in next sections), we obtain
\begin{eqnarray}
2\Lambda_{\rm eff}+\beta R_c^2=0 \Rightarrow R_c=\sqrt{\frac{-2\Lambda_{\rm eff}}{\beta}}.
\end{eqnarray}
This is consistent with the one obtained in the dilaton representation of the action, i.e. Eq.(\ref{S2}). Consequently we deduce that two frames give the same black hole solutions formally. A more realistic equivalence should be considered when we consider holographic renormalization in next section Sec.\ref{Holo}. 

\section{Scalar perturbations}
\label{spt}
It will be very illustrative to find full spectrum of the dilation field with EoM (\ref{eomphi}) on the background metric (\ref{ads2}). In the absence of matter fields $J=T_{\mu\nu}=0$, the trace of the Eq.(\ref{eomg}) simplifies to the following form,
\begin{eqnarray}
\nabla_{\mu}\nabla^{\mu}\phi+\theta\phi\lambda^{\mu}\nabla_{\mu}\phi-\beta\phi^2+\frac{\mu}{2}=0.
\end{eqnarray}
Note that here $\lambda=(\lambda^{0},\lambda^1)$. We suppose that $\phi(t,r)=\phi_c+\delta\phi(t,r) $ is an approximate solution. Using linear perturbation theory, here we assume that $\phi(t,r)$ depend on both time coordinate $t$ and radial coordinates $r$. Substituting the expression of $\phi(t,r)=\phi_c+\delta\phi(t,r) $  into (\ref{eomphi}), we obtain a differential equation for
 $\delta\phi(t,r)$,
\begin{eqnarray}\label{pert1}
r^2\delta\phi(t,r)''+\Big(2r+\phi_c\theta\lambda^1   \Big)\delta\phi(t,r)'+\partial_t\Big(2\delta\phi(t,r)'+\phi_c\theta \lambda^{0}\delta\phi(t,r)\Big)-2\phi_c\beta \delta\phi(t,r)=0.
\end{eqnarray}
We can substitute a Fourier decomposed form for field perturbation into (\ref{pert1}) such that
\begin{eqnarray}
\delta\phi(t,r)=\frac{1}{\sqrt{2\pi}}\int_{-\infty}^{\infty}e^{-i\omega t}\delta\tilde{\phi}(\omega,r)d\omega
\end{eqnarray} 
Note that we could use Fourier mode decomposition because the metric background Eq.(\ref{ads2}) is invariant under time boost $t\to t+c$ and it shows that the planar mode eigen-functions $e^{-i\omega t}$ still remain an eigen-function for the translational symmetry operator $\frac{\partial}{\partial t}$. This is due to the gauge invariant in two dimensions where the number of conformal copies is basically infinite. Note that the normalization  factor $\frac{1}{\sqrt{2\pi}}$ is chosen symmetrically. Hence one can use an unnormalized basis function instead. 

The general solution of (\ref{pert1}) is an arbitrary linear combination of the Bessel functions $J_{n}$, or equivalently, when $n\equiv \hat{n}\in\mathcal{C}$ is not an integer, of $I_{\hat{n}},K_{\hat{n}}$: 
\begin{eqnarray}
&&\delta\tilde{\phi}(\omega,r)=\frac{{{\rm e}^{\xi}}
}{\sqrt{r}}\Big(C(\omega)I _{\hat{n}}(\xi
)+D(\omega) K _{\hat{n}}(\xi
)\Big).
\end{eqnarray}
Here $C(\omega),D(\omega)$ are Fourier spectral amplitudes, and  
\begin{eqnarray}
&&\xi=\,{\frac {\lambda^{{1}}\phi_{{c}}(\Delta-\eta)  
-2\,i\omega}{2r}},\\&&
\hat{n}=\frac{1}{2}\,\sqrt {1+4\,\phi_{{c}}\mu+\phi_{{c}}(4
\,i\omega\,\lambda^{{0}}-8)(\Delta-\eta)}.
\end{eqnarray}
The modified Bessel functions of imaginary order are defined as 
\begin{eqnarray}
I_{\nu}(x)&=&e^{-i\nu \pi/2}J_{\nu}(xe^{i\pi/2}),\\
K_{\nu}&=&\frac{\pi}{2}\frac{I_{-\nu}(x)-I_{\nu}(x)}{\sin(\nu \pi)},
\end{eqnarray}
where $J_{\nu}(x)$ is the Bessel function. The asymptotic theory of the Bessel functions of imaginary order was developed in \cite{PRSA}. In our case at $r\to\infty$, we have
\begin{eqnarray}
I_{\nu}(x)&\approx&\frac{1}{(\nu)!}\big(\frac{x}{2}
\big)^{\nu}, x\ll1,\\
I_{-\nu}(x)&\approx&\frac{1}{(-\nu)!}\big(\frac{x}{2}
\big)^{-\nu}, x\ll1.
\end{eqnarray}
Consequently we have,
\begin{eqnarray}
K_{\nu}(x)\approx-\frac{\pi}{2(-\nu)!\sin(\nu \pi)}\big(\frac{x}{2}
\big)^{-\nu}.
\end{eqnarray}
Then the solution $\delta\tilde{\phi}(\omega,r)$ has the following form:
\begin{eqnarray}
\delta\tilde{\phi}(\omega,r)&\approx &-\frac{D(\omega)}{\sqrt{r}}\frac{\pi}{2(-\hat{n})!\sin(\hat{n} \pi)}\big(\frac {\Omega}{4r}
\big)^{-\hat{n}},\\
\Omega &=&\lambda^{{1}}\phi_{{c}}(\Delta-\eta)
-2\,i\omega.
\end{eqnarray}
Using inverse Fourier transform, the above equation implies
\begin{eqnarray}\label{delta}
\delta\tilde{\phi}(t,r)\approx  r^{\Re{\hat{n}(\omega=0)}-1/2}\cos(\Im{\hat{n}(\omega=0)}\ln r+\psi),
\end{eqnarray}
where $\Re,\Im$ are real and imaginary parts of $\hat{n}(\omega=0)\in\mathcal{C}$, $\psi\in\mathcal{R}$ is a phase factor.  Because Eq.(\ref{delta}) is the first order approximated solution for perturbations we observe that if $\Re{\hat{n}(\omega=0)}<1/2$, the $AdS_2$ is stable under scalar perturbations.

\section{Near-AdS$_2$ perturbations }
\label{adspt}
The AdS$_2$ dilatonic theories are widely investigated in the context of the near horizon geometries of extremal charged higher dimensional black objects \cite{Porfyriadis:2018jlw} as well as reduced lower dimensional bulk theories in reduction of higher order gravity theories beyond classical regimes \cite{Kolekar:2018sba}. For example in Ref.\cite{Porfyriadis:2018jlw}, the author showed that the geometry of  near horizon  of a near-extreme Reissner-Nordstrom (RN) black hole is near AdS$_2 \times S^2$. The geometry for near AdS$_2$ metric as a probe dilaton bulk black hole is given as follows:
\begin{eqnarray}
\frac{g}{L^2}=-r(r+2k)dt^2+\frac{dr^2}{r(r+2k)}, \ \ r\sim k\ll 1,\ \ 
 \phi=L(r+k),
 \label{nearads2}
\end{eqnarray}
In the approximated regime $r\sim k^p\ll 1,0<p<1$, the near AdS$_2$ metric Eq.(\ref{nearads2}) reduces to 
\begin{eqnarray}
\frac{g}{L^2}=-r^2dt^2+\frac{dr^2}{r^2}, \phi=Lr=\phi_b.
 \label{exactads2}
\end{eqnarray}
Note that both metrics (\ref{nearads2},\ref{exactads2}) solves EoMs presented in Eqs.(\ref{eomphi},\ref{eomg}). Regarding both metrics, we stress here that both of them are locally diffeomorphic. The main difference of them is the patches in which they cover on a Penrose diagram \cite{Kolekar:2018sba}. To investigate tensor perturbations, it is adequate to rewrite AdS$_2$ metric (\ref{exactads2}) in the following form in conformal gauge and using the lightcone coordinates $x^{\pm}=t\pm \frac{1}{r}$ being transformed to the following form 
\begin{eqnarray}
\frac{g}{L^2}=e^{2\omega_b}\Big(-dx^{+}dx^{-}
\Big),\ \ \omega_b=\ln r, \phi=\phi_b.
 \label{exactads23}
\end{eqnarray}
where $b$ refers to the background. The perturbations for  the metric and the dilaton Eqs.(\ref{exactads23}) are considered as 
\begin{eqnarray}
\omega=\omega_b+\delta\omega(x^{+},x^{-}),\phi=\phi_b+\delta\phi(x^{+},x^{-}).
\end{eqnarray}
Applying this to the EoMs given in Eqs.(\ref{eomphi},\ref{eomg}) we get
\begin{eqnarray}
&&\partial_{+}\partial_{-}\delta\phi(x^{+},x^{-})+\frac{2\delta\phi(x^{+},x^{-})}{(x^{+}-x^{-})^2}=0,\label{per1}\\&&
\partial_{+}\partial_{-}\delta\omega(x^{+},x^{-})+\frac{1}{(x^{+}-x^{-})^2}(2\delta\omega(x^{+},x^{-})-a\delta\phi(x^{+},x^{-}))=0,
\label{per2}
\end{eqnarray}
where $a$ is a constant. In the case of $a=0$, we can simply solve Eqs.(\ref{per1}) and (\ref{per2}) to obtain 
\begin{eqnarray}
\delta\phi(x^{+},x^{-})= \frac{c_{1}}{x^{+}-x^{-}},\quad  \delta\omega(x^{+},x^{-})= c_{2}(x^{+}-x^{-})^{2},
\label{tri}
\end{eqnarray}
where $c_{1}$ and $c_{2}$ are integration constants. In term of the original coordinates $(t,r)$, we have
\begin{eqnarray}
\delta\phi(x^{+},x^{-})=\frac{c_{1}}{x^{+}-x^{-}} = \frac{c_{1}r}{2},\quad  \delta\omega(x^{+},x^{-})= c_{2}(x^{+}-x^{-})^{2} = \frac{4c_{2}}{r^{2}}.
\label{tri}
\end{eqnarray}
Since
\begin{eqnarray}
\partial_{+}=\partial_{t}\frac{\partial t}{\partial x^{+}}+\partial_{r}\frac{\partial r}{\partial x^{+}}&=&\partial_{t}-r^{2}\partial_{r},\\ \partial_{-}=\partial_{t}\frac{\partial t}{\partial x^{-}}+\partial_{r}\frac{\partial r}{\partial x^{-}}&=&\partial_{t}+r^{2}\partial_{r},
\label{tri}
\end{eqnarray}
we find
\begin{eqnarray}
\partial_{+}\partial_{-}=\partial_{tt}-r^{2}\partial_{r}\left(r^{2}\partial_{r}\right).
\label{triad1}
\end{eqnarray}
Using the separation method of $\delta\phi(r,t)=R(r)T(t)$, therefore, Eq.(\ref{per1}) becomes
\begin{eqnarray}
\frac{1}{T(t)}\frac{d^{2}T(t)}{dt^{2}} =\frac{r^{2}}{R(r)}\frac{d}{dr}\left(r^{2}\frac{dR(r)}{dr}\right)-\frac{r^{2}}{2} ={\rm cont.}=-k^{2}.
\end{eqnarray}
Hence the solutions of $T(t)$ and $R(r)$ read
\begin{eqnarray}
T(t)=T_0 \sin(kt+\theta_{0}),\,\,R(r)=\sqrt{\frac{k}{2r}}\left[c_2  J_{-\sqrt{3}/2}\Big(\frac{k}{r}\Big)+c_3 J_{\sqrt{3}/2}\Big(\frac{k}{r}\Big)\right],
\label{genphi}
\end{eqnarray}
with $T_{0},\,\theta_{0},\,c_{2}$ and $c_{3}$ being constants. Therefore the general solutions take the form
\begin{eqnarray}
\delta\phi(r,t)=T_0 \sqrt{\frac{k}{2r}}\sin(kt+\theta_{0})\left[c_2  J_{-\sqrt{3}/2}\Big(\frac{k}{r}\Big)+c_3 J_{\sqrt{3}/2}\Big(\frac{k}{r}\Big)\right].
\label{genphic}
\end{eqnarray}
From the last expression we learn that dilaton AdS bulk action perturbations have asymptotically flat demolishing property where at $r\to\infty$, 
\begin{eqnarray}
&&\lim_{r\to\infty} |\delta\phi(r,t)|=0.
\end{eqnarray}
Furthermore at large but finite values of r, 
\begin{eqnarray}
|\delta\phi(r,t)|\leq \frac{T_0}{r^n},\ \ n>1,
\end{eqnarray}
consequently dilaton perturbation looks like just a charged $U(1)$ field in flat spacetime. 

In the same manner, we can quantify the solution of Eq.(\ref{per2}) using the same method. Supposing that $\delta\omega(r,t)=S(r)Q(t)$, we have
\begin{eqnarray}
\frac{1}{Q(t)}\frac{d^{2}Q(t)}{dt^{2}} =\frac{r^{2}}{S(r)}\frac{d}{dr}\left(r^{2}\frac{dS(r)}{dr}\right)-\frac{r^{2}}{2}+\frac{ac_{1}r^{3}}{8} ={\rm cont.}={\bar c}
\label{eqome}
\end{eqnarray}
In case of ${\bar c}=0$, we find for Eq.(\ref{eqome}) that
\begin{eqnarray}
Q(t)=c_3,\,\,S(r)=\frac{2 \sqrt{2} }{\sqrt{a} \sqrt{c_{1}} \sqrt{r}}\Bigg[c_{4}J_{\sqrt{3}}\Big(\frac{\sqrt{arc_1}}{\sqrt{2}}\Big)+c_5 J_{-\sqrt{3}}\Big(\frac{\sqrt{arc_1}}{\sqrt{2}}\Big)\Bigg],
\label{genome}
\end{eqnarray}
with $c_{3},\,c_{4}$ and $c_{5}$ being constants. Hence the general solutions read 
\begin{eqnarray}
\delta\omega(r,t)=\frac{c_32 \sqrt{2} }{\sqrt{ac_{1}r}}\Bigg[c_{4}J_{\sqrt{3}}\Big(\frac{\sqrt{arc_1}}{\sqrt{2}}\Big)+c_5 J_{-\sqrt{3}}\Big(\frac{\sqrt{arc_1}}{\sqrt{2}}\Big)\Bigg],
\label{genomec}
\end{eqnarray}
Notice that in this particular case the solution is time-independent. At the AdS boundary , when $r\to\infty$, we clearly observe that 
\begin{eqnarray}
&&\lim_{r\to\infty} |\delta\omega(r,t)|=0.
\end{eqnarray}
The metric of the AdS dilaton under perturbations "freezes". Under perturbations the shape of the AdS throat doesn't change. 

\section{Holographic renormalization}
\label{Holo}
Let us discuss now the conformal symmetry of dual model. Our results are very much inspired by the methods developed recently in \cite{prl} as an attempt to find  a simple dilaton toy model for Sachdev-Ye-Kitaev (SYK) model \cite{Sachdev1993,Kitaev2015}. The SYK models are nicely proposed as a solvable class of the quantum mechanical systems, where the interaction term is considered as a random parameter and very recently it has been suggested to have a gravity dual in two dimensional $AdS_2$ backgrounds.
In two dimensions the number of conformal symmetry generators are infinity and consequently the corresponding Hilbert space is infinite dimensional space. The reason is that  in the special case of $\mbox{dim} = 2$, the conformal Killing equation is reduced to  the Cauchy-Riemann equations. Thus in lower dimensional quantum systems, all holomorphic functions  are solutions for killing equation and they generate conformal coordinate transformations. It is obviously shown that in this case we have an infinite number of generators and we can find an infinite number of associated conserved charges in quantum  theory.

Because of this gauge freedom, we are free to take any type of time parametrization in the form of a conformal map like 
$t\to w,\ \ t = t(w)$. Such different parametrization changes metric under a certain class of conformal transformations and are characterized by 
a Weyl rescaling of the metric in the form 
$h_{\mu\nu}\to e^{2\Omega}
h_{\mu\nu}$. Such transformation is an isometric map and consequently it preserves the length between two points in the space time manifold. In gauge/gravity duality picture, conformal symmetry corresponds to
diffeomorphisms and it fixes the boundary metric. It is easy to show that under a  general conformal transformation $t\to t(w)$, and the AdS$_2$ vacuum (\ref{ads2}) transforms to the following form:
\begin{eqnarray}
&&\phi=\phi_c,\ \  g = -(r^2 + 2Sch(w))dw^2 + 2dwdr \label{adsbh}.
\end{eqnarray}
Here  $Sch(w)=\frac{t'''(w)}{t'(w)}-\frac{3}{2}\big(\frac{t''(w)}{t'(w)}\big)^2$  is the Schwarzian derivative \cite{prl}. 

Let us study a holographic renormalization group
(RG) flow which brings to an end in an AdS$_2$ throat. First we turn-off matter fields $T_{\mu\nu}=J=0$, and we attempt to connect smoothly the AdS$_2$ metric (\ref{ads2}) near-horizon metric (\ref{adsbh}) to an RG
flow at  enough large $r$ \cite{prl}. First of all we need to write Einstein equations given in (\ref{eomg}) for metric (\ref{adsbh}). If we linearize Einstein equations in a perturbative scheme, it is straightforward to show that the system of EOMs for $\phi$ and metric have the following solutions:
\begin{eqnarray}
&&\phi\approx \phi_c+l(rf_1(w)+f_2(w))+\mathcal{O}((lr)^2)\label{perphi},\\
&&
g= -(r^2 + 2 Sch(w))dw^2 + 2dwdr+\mathcal{O}((lr)^2),\label{perg}
\end{eqnarray}
where $f_i(w)$ are considered to be time perturbations and the series expansions are valid till only when $lr\ll1$ and $l$ is an IR cutoff. By solving  $r-r,r-w$ components of Einstein equation (\ref{eomg}), we can find that $f_1(w)=1,f_2(w)=0$. The enough and sufficient condition to satisfy both Einstein and scalar field equations is that the Schwarzian function $Sch(w)$ must satisfy the following differential equation, called as continuity equation,
\begin{eqnarray}
l\frac{d\ln Sch(w)}{dw}\approx \frac{(\Delta+\Delta^{*}-2\eta)(2+\sqrt{\mu})}{2\sqrt{\mu}}.\label{continuty}
\end{eqnarray}
Because we need the RG flow to bring to an end in an AdS$_2$ black hole, we need to have vanishing terms in the right hand side of Eq.(\ref{continuty}). It  gives us the following constructive relation between Kitaev parameters:
\begin{eqnarray}
\eta=\frac{\Delta+\Delta^{*}}{2}.
\end{eqnarray}
Note that here we can also interpret (\ref{continuty}) as an extended version  of Eq.(20) in Ref.\cite{prl}, when we realized model by the same quantum parameters. We can rewrite (\ref{continuty}) in the following form:
\begin{eqnarray}
l\frac{dSch(w)}{dw}=-T_{ww}, \label{tww}
\end{eqnarray}
where $T_{ww}\equiv  -\frac{(\Delta+\Delta^{*}-2\eta)(2+\sqrt{\mu})}{2\sqrt{\mu}}Sch(w)$. We can convert Eq.(\ref{tww}) into an equation in the boundary
quantum mechanics. Using holographic renormalization in order $\mathcal{O}(l)$ we find that  the boundary energy is defined as follows:
\begin{eqnarray}
E=-\frac{l}{\kappa^2}\Big(\ln Sch(w)-\frac{\theta\kappa^2 (2+\sqrt{\mu})}{\sqrt{\mu}}w\Big).
\end{eqnarray}
As a result we can write an effective action for dilaton gravity dual to Kitaev model, which is valid only in the vicinity of the AdS throat
\begin{eqnarray}
S_{\rm eff}\approx W_{CQM}[\beta;t(w)]-\frac{lw^2}{\kappa^2}\Big(w^{-1}\ln Sch(w)-\frac{(\Delta+\Delta^{*}-2\eta)(2+\sqrt{\mu})}{4\sqrt{\mu}}-\frac{ \kappa^2(2+\sqrt{\mu})\theta}{\sqrt{\mu}}
\Big).\label{Seff}
\end{eqnarray}
Actually here $W_{CQM}$ is the “generating functional”. 
The above expression defines an effective quantum action for our Kitaev theory. 

\section{Ryu-Takayanagi holographic entanglement entropy in the Kitaev model}
\label{entang}
An entanglement entropy (EE) for Kitaev model is derived in Ref.\cite{PRB}. It is illustrative to examine our bulk toy model to find EE using holographic principle. We analyze the HEE for the toy model of Kitaev model presented in Eq.(\ref{S2}). We thus analyze the HEE Eq.(\ref{hee}) of a two-dimensional AdS black hole Eq.(\ref{perg}) and relate it to the EE of a boundary quantum theory theory. Because the metric form given in Eq.(\ref{perg}) is time dependent, we need to analyze the time dependent geometries using the concept of time dependent HEE. One can generalize  the above conjecture for the HEE to time-dependent AdS  bulk geometries \cite{timedephee}. Very recently HEE and phase transition of a 2-dimensional holographic superconductor have been investigated in Ref. \cite{Momeni:2015iea}  and furthermore we developed a systematic way to calculate HEE and fidelity susceptibility in time dependent backgrounds using AdS/CFT \cite{Momeni:2016ira}.

In the case of time-dependent bulk theories (dual to out of the equilibrium CMT models ), the HEE in the time-dependent geometries is very different from the static cases since their dual CMT systems are different (one is at equilibrium and the other is out of the equilibrium). In the time-dependent geometries like our background, it is not possible to foliate a bulk time-dependent geometry by a preferred time slicing. It is possible to foliate a time-dependent asymptotically AdS geometry by zero mean curvature slicing. Thus, it is only possible to take slices of the bulk geometry with vanishing trace of extrinsic curvature. This corresponds to taking the spacelike slices with maximal area through the bulk, anchored at the boundary. This covariant foliation reduces to the constant time foliation for static geometries. Thus, a co-dimension one spacelike foliation of time-dependent asymptotically AdS geometry can be performed, and on such a spacelike slice the metric is spacelike. As a result, a co-dimension two minimal surface can be defined on such a spacelike slice. So, in this formalism, first a maximal spacelike slice of the bulk geometry is obtained though the mean curvature slicing, and then a minimal surface $\gamma _{A}$ for the entangled region $A$ is constructed on this spacelike slice.

Let us start finding minimal surface for metric  Eq.(\ref{perg}). This bulk geometry is the Vaidya spacetime
\begin{eqnarray}
\lim_{u\to 0}  ds^2=\frac{1}{u^2} \big(-F(w,u)du^2-2dwdu\big)+\mathcal{O}(u^2)\label{g2}
\end{eqnarray}
where $F(w,u)=1+2u^2Sch(w)$ is a  function of the ingoing Eddington-Finkelstein time coordinate $w$, and $u$ is introduced as Poincar\'e coordinate. Note that the minimal area is really the minimal length. It is worth noting that when $u=0$, we obtain the AdS boundary. Although it is not possible to define a physical temperature for a time-dependent backgrounds because there is no time Killing vector  it has been proved that the time- dependence backgrounds can be analyzed as a perturbation of static AdS geometry, and  we will  use it  for computing the time-dependence HEE for our toy model. We will suppose that the entangled region is defined as a strip geometry such that its width is $2l$ in the $u$ direction. Now because of the symmetries of the surface, $w = w(u)$ is only a function of $u$, and the surface $\gamma_{Aw}$ will be characterized by the embedding
\begin{eqnarray}
\gamma_{Aw}=\{w = w(u)\}.
\end{eqnarray}
The area functional as a function of the surface, $w = w(u)$ is only a function of $u$. We observe that the possible extremal surface will be smoothly extended into the bulk, where the center of the strip is located at $u = 0$ and the AdS boundary in Poincare coordinate corresponds to $ u = 0$. In holographic computation of the EE, the main contribution to the minimal area function and extremal curve belong to the AdS boundary for $u\to 0$. This is the main reason why the computation the minimal surface here is also important only for the nearly boundary region. Basically we know that in the AdS/CFT, the quantum boundary theory is in thermal equilibrium with the gravity sector in the bulk, and the junction region is AdS boundary section, i.e. $u=0$, and it satisfies the following relation on the boundary
\begin{eqnarray}
&&w(u=0)=w^{*},\ \ w'(u=0)=0\label{bcs}.
\end{eqnarray}
The turning point of the strip is located at $u=0$. At the boundary $u=0$, the physical time is $T=w+u$ and at $u = l$, we need to impose an auxiliary UV boundary conditions (BCs),
\begin{eqnarray}
&&w(u=l)=T-\epsilon.
\end{eqnarray}
Here we inserted  a cut-off $\epsilon$ to deal with the UV divergence at the AdS boundary. We can express the area of the minimal surface $\gamma_{Aw}$ as
\begin{eqnarray}
&&\mbox{Area}(\gamma_A)=\int_{-l}^{l}du\frac{\sqrt{-\big(1+2u^2Sch(w)+2w'(u)\big)}}{u}.\label{area}
\end{eqnarray}
We also fix the strip size to be \footnote{Note that the condition $1+2u^2Sch(w)+2w'(u)<(=)0$ for a non uniform $Sch(w)$ leads to a nonlinear extension of Bernoulli Equation for $w(u)$. }
 \begin{eqnarray}
 2l=\int_{\epsilon}^{w^{*}}\frac{dw}{w'(u(w))}\label{L}.
 \end{eqnarray}
Using the Euler-Lagrange (EL) equation imposed on Eq. (\ref{area}) we obtain,
\begin{eqnarray}
\frac{d}{du}\Big[\frac{1}{uQ}
\Big]&=&\frac{\partial}{\partial w}\Big[\frac{uSch'(w)}{Q}
\Big],\\
Q&=&\sqrt{-\big(1+2u^2Sch(w)+2w'(u)\big)}.\label{EL}
\end{eqnarray}
Using Eq.(\ref{continuty}) we deduce that 
\begin{eqnarray}
&&\frac{d}{du}\Big[\frac{1}{uQ}
\Big]=\frac{(\Delta+\Delta^{*}-2\eta)(2+\sqrt{\mu})}{2l\sqrt{\mu}}\frac{\partial}{\partial w}\Big[\frac{u}{Q}
\Big].\label{EL2}
\end{eqnarray}
Note that $\frac{\partial}{\partial w}(Q)=-\frac{u^2Sch'(w)}{Q} $. Consequently we obtain the ultimate form of EL Eq. as,
\begin{eqnarray}
&&w''+4\gamma u w'+8Sch(w)+\frac{4}{u^2}+4\gamma^2\frac{u^3}{Q^2}=0.\label{EL3}
\end{eqnarray}
where $\gamma=\frac{(\Delta+\Delta^{*}-2\eta)(2+\sqrt{\mu})}{2l\sqrt{\mu}}$ is a chaos parameter. This differential equation models hybrids chaotic system with both continuous and discrete dynamics. Three different phases exist \cite{PRB}: the zero quasi energy Majorana (MZM) phase
($0 < \frac{\mu}{\eta} < 2$), the trivial phase ($2 < \frac{\mu}{\eta} < 3$), and the Majorana $\pi$ (MPM) phase ($3 < \frac{\mu}{\eta} < 4$). This chaotic differential equation for minimal surfaces explains the possible reason to have chaos in AdS$_2$, in a different terminology in comparison to the Ref.\cite{prl} and we will investigate it in details for this work.

By the way we focus only on the case of constant  Schwarzian obtained using a conformal map $t(w):w\rightarrow \tanh(\pi w T)$, where $Sch(w)=-2\pi^2 T^2$\cite{prl}. We solve Eq.(\ref{EL3})  when $\gamma=0$ to obtain
\begin{eqnarray}
w_0(u) =  w^{*} + 8 \pi^2 T^2 u^2 + 4( \ln\frac{u}{\epsilon}-\frac{u}{\epsilon}),\ \ \epsilon\to 0.
\end{eqnarray}
Note that the solution is subjected to BCs, (\ref{bcs}). Using perturbation theory the first approximated solution for Eq.(\ref{EL3}) is given by:
\begin{eqnarray}
&&w(u) = w_0(u)+\gamma w_{1}(u).
\end{eqnarray}
A straight forward calculation shows that
\begin{eqnarray}
&& w_{1}(u)=-(\frac{16}{3}\pi^2T^2)u^4+(\frac{8}{3\epsilon})u^3-8u^2+w^{*}
\end{eqnarray}
Note that the BCs are imposed also on the first perturbated solution as well as zeroth order function to fulfills all requirements.

Using this solution, the area defined in Eq. (\ref{area}) can be now expressed as,
\begin{eqnarray}
&&\mbox{Area}(\gamma_A)=\int_{-l}^{+l}du\frac{Q_0}{u}+\gamma\int_{-l}^{+l}\frac{du}{u}\frac{\partial Q}{\partial w'}|_{w'=w_0'}\,,\label{integral2}
\end{eqnarray}
where 
\begin{eqnarray}
Q_0 &=&\sqrt{-\big(1-4\pi^2T^2u^2+2w_0'(u)\big)},\\
 \frac{\partial Q}{\partial w'}&=&-\frac{w'}{Q}|_{w'=w'_0}=-\frac{w_0'}{Q_0}.
\end{eqnarray}
It is interesting to be noted that usually there is (are) UV divergence term(s) in HEE, and so we need to use 
a regularization method to improve these divergence terms. Thus, for a deformed geometry, we define the area  as,
\begin{eqnarray}
\mathcal{A}(\gamma_A) = \mathcal{A}_{D}(\gamma_A) - \mathcal{A}_{AdS}(\gamma_A), 
\end{eqnarray}
where  $ \mathcal{A}_{D}(\gamma_{A}) $ is the defined in deformed geometry (for example the geometry of an AdS  black hole with $w(u)$), and $ \mathcal{A}_{AdS}(\gamma_A)$
is defined in the background $AdS$ spacetime where $w(u)=w_0(u)$. Thus, we define the HEE for a deformed geometry by subtracting the 
term coming from the unperturbated $AdS$ throat. This only leaves a finite part. 
We used this finite part in our former papers Refs. \cite{Momeniplb1},\cite{Momeniplb2}, and call it the HEE. In our case (\ref{integral2}), the first integral is the divergent part and consequently we define the holographic form of $S_{EE}$ as the following integral:
\begin{eqnarray}
S_{\rm HEE}=-\frac{\gamma}{2\pi TG}\int_{-l}^{+l}\frac{4\pi^2T^2u^2+1}{\sqrt{u^{3}(u-u_1)(u-u_2)(u-u_3)}}du\label{hee1}.
\end{eqnarray}
which is an elliptic integral for $l\to\infty$ and $u_{1},u_{2},u_{3}$ are roots of the following cubic equation
\begin{eqnarray}
&&4\pi^2T^2u^3-32\pi^2T^2u^2-u-8=0\label{cubic}.
\end{eqnarray}
These roots are presented as following:
\begin{eqnarray}
&&u_1=\frac{\delta }{6 \pi  T^2}-\frac{-1024 \pi ^4 T^4-12 \pi ^2 T^2}{24 \pi
	^3 \delta  T^2}+\frac{8}{3},
\\&&
u_2=-\frac{\left(1-i \sqrt{3}\right) \delta }{12 \pi  T^2}+\frac{\left(1+i
	\sqrt{3}\right) \left(-1024 \pi ^4 T^4-12 \pi ^2 T^2\right)}{48 \pi ^3
	\delta  T^2}+\frac{8}{3},
\\&&
u_3=-\frac{\left(1+i \sqrt{3}\right) \delta }{12 \pi  T^2}+\frac{\left(1-i
	\sqrt{3}\right) \left(-1024 \pi ^4 T^4-12 \pi ^2 T^2\right)}{48 \pi ^3
	\delta  T^2}+\frac{8}{3},\\&&
\delta=\sqrt[3]{4096 \pi ^3 T^6+288 \pi  T^4+3 \sqrt{3} \sqrt{65536 \pi ^4
		T^{10}+2816 \pi ^2 T^8-T^6}}.
\end{eqnarray}
Equation (\ref{cubic}) shows that only the $u_1\in\mathcal{R}$ is analytic. It can be shown that $u_2,u_3\in\mathcal{C}$. If we fix the entangled length as $l>0$, we can integrate and write down final expression for HEE,
\begin{eqnarray}
S_{\rm HEE}=-\frac{\gamma}{\pi TG} \Bigg[{\cal E}_{1}\,{\cal E}_{2} \Bigg({\cal E}_{3}+ {\cal E}_{4} + {\cal E}_{5} \Bigg)\sqrt {\frac {u (u_1 - u_3)} {u_3 (u_1 -u)}} + u_1\Bigg]^{u=l}_{u=-l},\label{hee}
\end{eqnarray}
where
\begin{eqnarray}
\nonumber {\cal E}_{1}&=&\frac{ u (u -u_2) (u_3 - u)}{u_1 u_2\sqrt {u^3 (-(u_1 - u)) (u - u_2) (u - u_3)}}\,,\\\nonumber {\cal E}_{2}&=&\frac{1}{(u_2 -u_3)\sqrt {-\frac {(u_1 - u_2) (u_1 - u_3) (u_2 - u) (u_3-u)} {(u_1 - u)^2 (u_2 - u_3)^2}}}\,,\\\nonumber {\cal E}_{3} &=& u_2\left (a u_1^2 + 1 \right) F\left (\sin^{-1}\left (\sqrt {\frac {(u_1 -u_2) (u_3 - u)} {(u_3 - u_2) (u_1 - u)}} \right) | \frac {u_2 (u_3 - u_2)} {(u_1 - u_2) u_3} \right)\,,\\\nonumber {\cal E}_{4}&=&a u_1 u_2 (u_3 - u_1)\Pi\left (\frac {u_3 - u_2} {u_1 - u_2};\sin^{-1}\left (\sqrt {\frac {(u_1 - u_2) (u_3 -u)} {(u_3 - u_2) (u_1 - u)}} \right) | \frac {u_1 (u_3 -u_2)} {(u_1 - u_2) u_3} \right)\,,\\\nonumber {\cal E}_{5}&=&(u_1 -u_2) E\left (\sin^{-1}\left (\sqrt {\frac {(u_1 - u_2) (u_3 -u)} {(u_3 - u_2) (u_1 - u)}} \right) | \frac {u_1 (u_3 -u_2)} {(u_1 - u_2) u_3} \right),
\end{eqnarray}
where $F,\,\Pi,\,E$ are Jacobi's elliptic functions. We numerically found that the exact results for the EE in the Kitaev model as a function  for $|\Delta,\eta |<3$  respectively presents a peak at $\mu=2\eta$ around the critical point of the Kitaev model. Note that the location of the parameter where the EE is maximum is different for different parameters.

We further study the variation of the regularized HEE in the dual  model Eq. (\ref{hee}). The analysis showed that  the  peak value of the HEE $|S_{HEE}|$ for different $T$ is in the form of the parabolic function being, $bT^2$. Thus the critical HEE of the dual model to Kitaev, is $S_{\rm HEE}\approx T^2$, which is close to the exact result (\ref{hee}) for $\mu=2\eta$ which agrees with the exact result of the correlation EE of the Kitaev model~\cite{PRB}.

\section{Fold bifurcation minimal AdS$_2$ surfaces}
\label{fold}
It has been shown that in the generalized Kitaev model, called Sachdev-Ye-Kitaev (SYK) \cite{Sachdev1993,Kitaev2015}, the hydrodynamic universally describes  large $N$ systems with an emergent conformal invariance, and consequently any such system will be maximally chaotic \cite{prl}. As we know, the chaotic structure in phase space are classified as follows:
\begin{itemize}
	\item Limit cycles are the ellipse-like paths with frequencies greater than the natural frequency of the system.
	\item Strange attractors: Those are highly sensitive to initial conditions.
	\item Casual attractors: The late time behavior of the dynamical system remains independent from the initial conditions which are imposed.
	 \item Chaotic paths: We observe rapidly fluctuations in the phase portrait.
\end{itemize}

In this section, we show that the chaotic behavior can be generated because of chaotic structure of the entanglement surfaces given in Eq.(\ref{EL3}) near criticality $\mu=2\eta$. Let us investigate the entangled surfaces equation of motion, when we consider $\gamma$ as a critical chaotic parameter, note that we can safely ignore the $w''$ if we consider the temperature enough high, consequently the equation becomes,
\begin{eqnarray}
\gamma u w'+u^{-2}-\frac{u^3 \gamma^2}{1-8u^2\pi^2 T^2+2w'}=4\pi^2T^2\label{EL31}
\end{eqnarray}
This eq. can by inverted to find $w'$,
\begin{eqnarray}\label{dsys}
w'=\frac{\mathcal{A}^{1/2}+8\gamma \pi^2 u^5-\gamma u^3+8\pi^2 u^2 T^2-2}{4\gamma u^3},
\end{eqnarray}
where we have defined, 
\begin{eqnarray}
\mathcal{A}&=&64\gamma^2\pi^4u^{10}+(8\gamma^3-16\gamma^2\pi^2)u^8-128T^2\gamma\pi^4u^7+\gamma^2u^6\\&&\nonumber+(16T^2+32)\gamma\pi^2 u^5+64 T^4\pi^4 u^4-4\gamma u^3-32T^2\pi^2 u^2+4,
\end{eqnarray}
where we have opted the positive root of $w'$. We numerically plot the vector field using \textsc{dfield } open source code for different values of $\gamma$ and for $T^{-1}=2\pi,\gamma=1$. We run the code using a sample of $(u_{min},u_{max}),(w_{min},w_{max})$, for initial condition $w(u=0)=w^{*}$.   
\begin{figure}[ph]
  \begin{center}
   \includegraphics[width=0.7\textwidth]{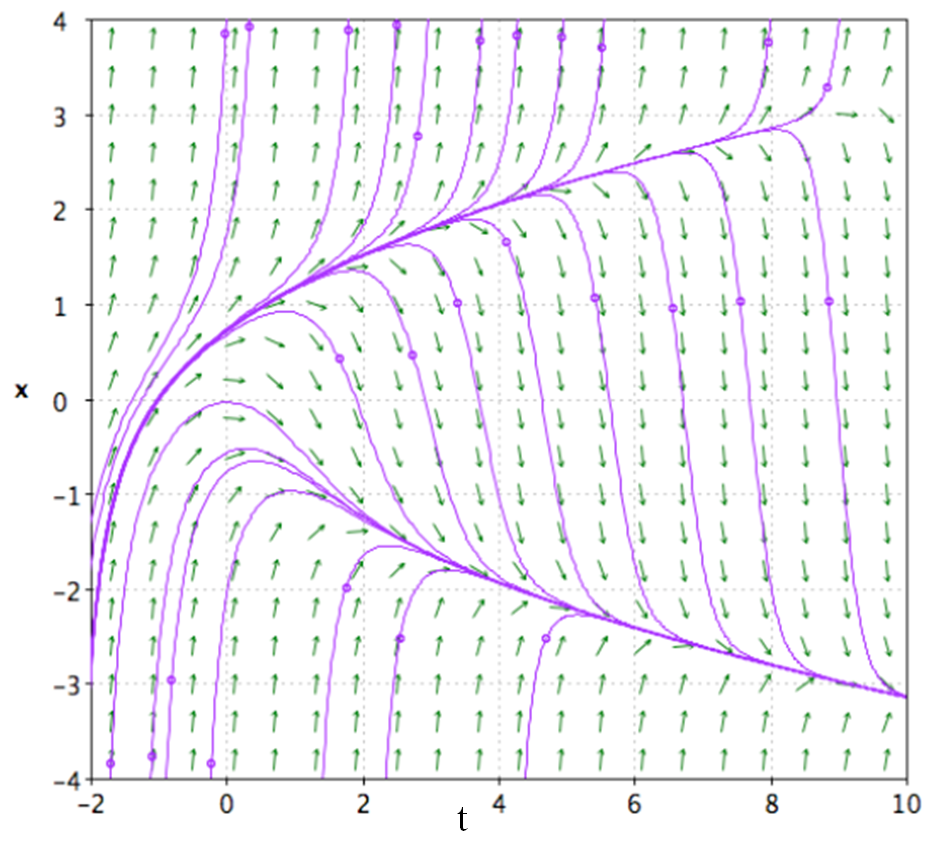}
  \end{center}
  \caption{Vector field description and the numerical solution for entangled surface $w(u)$ in the first approximation regime. Note that the horizontal axis $t$ is equal to $u$ and the vertical axis $x$ is $w(u)$. The coordinate $u$ plays the role of the time in the dynamical system given in Eq.(\ref{dsys}).}
  \label{wpw}
\end{figure}
In Fig.(\ref{wpw}), the green vectors show the direction of the gradient $w'(u)$ at each point while the blue curves show the numerical solution for the entangled surface. The qualitative form of the entangled surface dramatically changed by considering different values of the control parameter $\gamma$. There is a fold bifurcation. Furthermore , we note that there are three classes of the entangled surfaces. For a long enough time interval, there is an asymptotically convergence solution, the one on the bottom, which is converged to the fixed point $w^{*}$. Note that the middle class of the entangled surfaces are also convergent to the same fixed point family. The upper branch of the entangled surfaces are monotonically divergent and it indicate that the other branch never tends to any fixed point. Briefly we summarize that the entangled surface undergoes a fold bifurcation form, from one fixed point to no fixed point. This qualitative analysis shows that the origin of the proposed chaotic behavior could be related to the fold bifurcated minimal surfaces in the AdS$_2$.   

Let us to consider bifurcation idea hidden behind the dynamical system Eq.(\ref{EL31}), where the entangled surface subjected a non linear highly sensitive to the control parameter $\gamma$. We use the techniques developed in Ref.\cite{chaos}. The fixed points of Eq.(\ref{EL31}) are found to be the roots of the equation $w'=0$, or $f(u_c,\gamma)=0$, where 
\begin{eqnarray}
(u_c^{-2}-4\pi^2T^2)(1-8u_c^2\pi^2 T^2)=u_c^3 \gamma^2 .\label{uc}
\end{eqnarray}
We plot  the equilibrium manifold $f(u_c,\gamma)=0,\ \ u_c=u_c(\gamma,T)$, for $T^{-1}=2\pi,\,\pi,\,\pi/2,\,3\pi/4,\,\pi/4$.
\begin{figure}[ph]
  \begin{center}
   \includegraphics[width=1.0\textwidth]{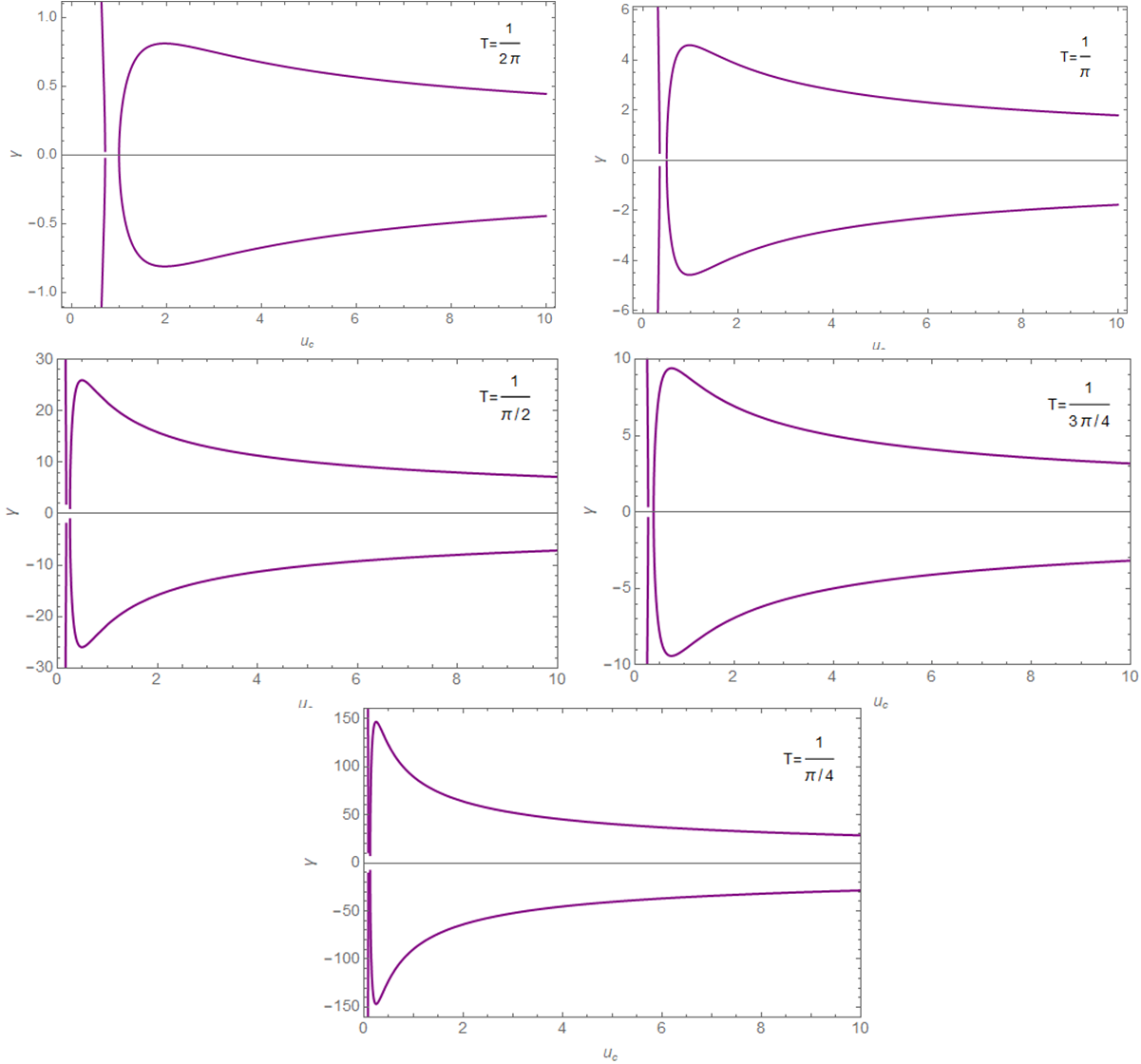}
  \end{center}
  \caption{The plots show the fixed point $u_c$ given in (\ref{uc}) versus the control parameter $\gamma$ for various $T^{-1}=2\pi,\,\pi,\,\pi/2,\,3\pi/4,\,\pi/4$. Notice that we observe bifurcations in our model.}
  \label{unif}
\end{figure}
\newpage
It clearly shows the fold bifurcation behaviors for varying values of $\gamma$. We illustrate the emergence of the fold bifurcation in the entangled surface parametrized as $w(u)$. Although we plotted the fixed point $u_c$ versus control parameter $\gamma $ in different ranges of temperature, we clearly observe a uniform pattern. Note that for $\gamma=0$, there is no any fixed point. It means that the long term behavior of the entangled surface (here $u\to 1$) does not approach any asymptotic value. When $\gamma\to 0$ still the system does not have any fixed point. When eventually the control parameter $\gamma\to 0^{\pm}$, for each value of $\gamma$ we have a unique $u_c$, the system enjoys a reflection symmetry $\gamma\to-\gamma$ and $u_c$ is an even function of $\gamma$ i.e. $u_c(\gamma)=u_c(-\gamma_c)$. As we have learned in the advanced dynamical systems language, the system undergoes a $0\to 1$  using the bifurcation diagram for unstable saddle node bifurcation $u_c$. This instability in the fixed points causes an unstable entangled surface $w(u)$.  

\section{Summary}
In conclusion, we have studied the possible Kitaev/AdS$_2$ duality between AdS$_2$ modified dilatonic action and Kitaev quantum wire, as a one dimensional lattice model for a superconductor. From the Kitaev/AdS$_2$ duality, we showed that the bulk toy model displays the same quantum phase transition on the critical points. Also, we used the holographic renormalization technique to study the effective quantum action. We wrote an effective action for dilaton gravity dual to Kitaev model, which is valid only in the vicinity of the AdS$_2$ throat.
 Moreover, we showed that the phase transition of the entanglement entropy can be related to the holographic entanglement entropy. Our approach directly showed that both the AdS$_2$ and the quantum model can be used to characterize the possible  quantum phase transition in  Kitaev model.

In a study of the hydrodynamical properties of the generalized Kitaev model, called Sachdev-Ye-Kitaev (SYK)~\cite{Sachdev1993,Kitaev2015}, it was found that the dual is maximally chaotic. However, only the Kitaev model is a two-body model and is not chaotic. It is just a fermi liquid where the single particle energies
are filled up to the Fermi surface to get the ground state energy. Also the $q=2$ SYK model is a standard Fermi liquid and is not chaotic. The $q\ge 4$ SYK models are completely different, they are chaotic, and are not a Fermi liquid. For example the zero temperature entropy is
extensive. Scalar and tensor perturbations have been investigated in details. The time-dependent part of the perturbation was an oscillatory model. We discovered that the dual gravity induces an effective and renormalizable quantum action. The entanglement entropy for bulk theory has been computed using extremal surfaces. We proved that these surfaces have a fold bifurcation regime of criticality. A thorough check showed that the entangled surface shape is very sensitive to the control parameter $\gamma$ which is defined in terms of the Kitaev parameters.

\section*{Acknowledgments}
D. Momeni work supported by the Internal Grant (IG/SCI/PHYS/20/07) provided by Sultan Qaboos University. P. Channuie acknowledged the Mid-Cereer Research Grant 2020 from National Research Council of Thailand under a contract No. NFS6400117.

\end{document}